\documentclass[prl, preprint, floatfix,superscriptaddress]{revtex4}
\usepackage{graphicx}
\usepackage{epstopdf}
\usepackage{color}
\usepackage{amsmath, amsfonts, amssymb, bm}
\begin{document}
\title{Amplifying ultraweak transitions in collective systems via quantum interference}
\author{Ni \surname{Cui}}
\email{ni.cui@mpi-hd.mpg.de}
\affiliation{Max-Planck-Institut f\"{u}r Kernphysik, Saupfercheckweg 1, D-69117 Heidelberg, Germany }
\affiliation{Siyuan Laboratory, Guangzhou Key Laboratory of Vacuum Coating Technologies and New Energy Materials, 
Department of Physics, Jinan University, Guangzhou 510632, China}
\affiliation{Guangdong Provincial Key Laboratory of Optical Fiber Sensing and Communications, Jinan University, Guangzhou 510632, China}
\author{Mihai A. \surname{Macovei}}
\email{macovei@phys.asm.md}
\affiliation{Max-Planck-Institut f\"{u}r Kernphysik, Saupfercheckweg 1, D-69117 Heidelberg, Germany }
\affiliation{Institute of Applied Physics, Academy of Sciences of Moldova, Academiei str. 5, MD-2028 Chi\c{s}in\u{a}u, Moldova}
\date{\today}
\begin{abstract}
We investigate laser-induced quantum interference phenomena in superradiance processes and in an ensemble of 
initially excited $\Lambda-$type closely packed three-level emitters. The lower doublet levels are pumped with a 
coherent laser field. Due to constructive quantum interference effects, the superradiance occurs on a much weaker 
atomic transition which is not the case in the absence of  the coherent driving. This result may be of visible relevance 
for enhancing ultraweak transitions in atomic or atomic-like systems, respectively, or for high-frequency lasing effects.
\end{abstract}
\maketitle
\noindent {\it Introduction:} Vacuum induced correlations among closely spaced quantum emitters forming an ensemble 
lead to significant changes in the quantum dynamics \cite{dicke,agarwal,gross,andre,Ficek,rew,gxl}. In this context, the 
superradiance - an already well-known phenomenon - emphasizes the fast decay of an initially excited cooperative 
system as well as an enhanced radiation intensity, respectively. An enormous amount of experimental and theoretical 
works were performed with respect to this issue and superradiance behaviors were found in a wide range of different 
systems and for various applications \cite{kaiser,garr,harr,carsten,nucl,svid,yelin,yavuz,paolo,solid}. The collective 
quantum dynamics can be manipulated by applying external coherent laser sources. Particularly, triggering of the 
superradiance phenomenon nicely occurs in a three-level Vee-type atomic ensemble when an external coherent field 
pumps one of the atomic transitions \cite{chk}. In a somehow related setup, superfluorescence without inversion was 
shown to occur as well \cite{koch}, see also \cite{wel}. One may anticipate cooperative effects in novel systems because 
x-ray free-electron lasers may accelerate the decay of a nuclear isomer \cite{adriana}. Furthermore, in a large ensemble 
of nuclei operating in the x-ray regime and resonantly coupled to a common cavity environment, two fundamentally different 
mechanisms related to cooperative emission and magnetically controlled anisotropy of the cavity vacuum have been responsible 
for fascinating effects mainly related to quantum interference phenomena \cite{jorg}. Actually, these effects originate 
from indistinguishability of the corresponding transition pathways \cite{Ficek,rew,jorg,pasp}. 

Interfering transition amplitudes can be used in principle to detect weak atomic interactions like measurements of 
magnetic dipole interactions, quadrupole interactions or weak atomic transitions occurring, for instance, because 
of parity violation effects  as well as to identify various nonlinear transition channels \cite{par,weak,weakk,min}. 
Furthermore, ultranarrow absorption lines due to electromagnetically induced transparency phenomenon were 
shown to be very useful for high-accuracy optical clocks \cite{santra}. Quite recently, superradiance on the 
millihertz linewidth strontium clock transition was shown to occur in \cite{sup_clock}. This was achieved with 
the help of an optical cavity which triggered the superradiance on the ultraweak transition. Somehow related, 
prospects for millihertz-linewidth lasers were suggested too in \cite{mill}.

Under these circumstances, we discuss here a setup where weak or ultraweak decaying transitions can be significantly 
enhanced in an initially excited ensemble of few-level collectively interacting $\Lambda-$ type atoms. Notably, the 
effect arises due to quantum interference phenomena among different decaying pathways which are induced by the 
coherent pumping of the two lower levels. The rapid time-evolution on a ultraslow atomic transition is determined by the 
fast decay rate on another transition of the $\Lambda$ sample. Moreover, the effect is of the cooperative nature and it is 
absent in excited single-atom systems or independent atomic ensembles, respectively. Particularly, (i) we have found 
that the superradiance on the ultraweak transition may take place when there are more atoms on the ground state than 
in the excited one; (ii) the superradiance peak occurs when the population of the excited level is trapped and almost 
constant during a short time which is distinct from the standard superradiance phenomenon where its time-dependent 
intensity relies on the fast population slope; (iii) quantum coherences induced by the coherent pumping are responsible 
for superradiant population transfer on ultraweak transition as well as among the lower sublevels during the superradiant 
burst. As possible applications of our results we suggest enhancing dipole-forbidden or any other ultraweak transitions, or
in quantum clocks atomic systems, respectively, as well as for high-frequency lasing.

\begin{figure}[t]
\begin{center}
\includegraphics[height=3.5cm]{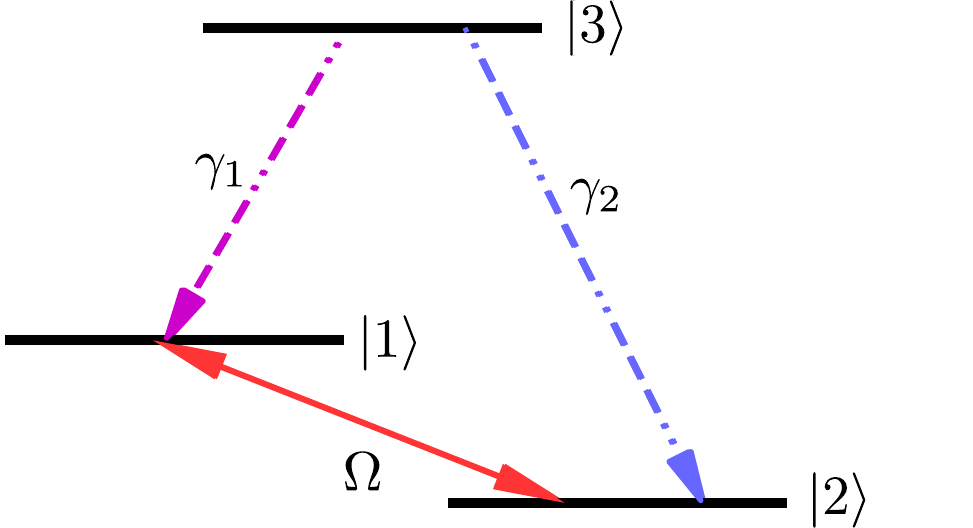}
\end{center}
\caption{\label{fig-1}(Color online) The energy levels of the $\Lambda-$type three-level system. $\gamma_{1}$ and $\gamma_{2}$,
 $\gamma_{1} \gg \gamma_{2}$, are the single-atom spontaneous decay rates on transitions $|3\rangle \to |1\rangle$ and 
$|3\rangle \to |2\rangle$, respectively. The coherent laser field drives the $|1\rangle \leftrightarrow |2\rangle$ transition with 
$\Omega$ being the corresponding Rabi frequency.}
\end{figure}

\noindent {\it Analytical framework:} We consider an initially excited ensemble of $N$ identical $\Lambda-$type three-level 
emitters each consisting of states $|1\rangle$, $|2\rangle$ and $|3\rangle$, as depicted in Figure~\ref{fig-1}. The lower 
doublet transition $|2\rangle \leftrightarrow |1\rangle$ is resonantly driven by a coherent laser field. The emitters can decay 
via a fast dipole-allowed $|3\rangle \leftrightarrow |1\rangle$ transition as well as through a slow or ultraslow 
$|3\rangle \leftrightarrow |2\rangle$ atomic transition, respectively, due to coupling with the environmental vacuum modes. 
The interparticle separations are of the order of relevant emission wavelengths of the system, or smaller, and in this way 
the atomic sample acquires a cooperative nature.

In the usual mean-field, Born-Markov and rotating-wave approximations, our model is described by the following master equation 
\cite{agarwal,gross,andre,Ficek,rew}
\begin{eqnarray}
\dot{\rho}(t) &+& i\Omega\sum_{j=1}^{N} \bigl[ S_{12}^{(j)} +S_{21}^{(j)},\rho \bigr]=-\sum_{j,l=1}^{N}\bigl \{ \gamma_{jl}^{(1)}[S_{31}^{(j)},S_{13}^{(l)}\rho]
\nonumber \\
&+& \gamma_{jl}^{(2)}[S_{32}^{(j)},S_{23}^{(l)}\rho] \bigr \}+ \text{H.c.}.\label{meq}
\end{eqnarray}
Here $\Omega$ is the corresponding Rabi frequency, while $\gamma_{jl}^{(s)}\equiv\gamma_{s}\left[\aleph_{jl}^{(s)} + i\Omega_{jl}^{(s)}\right]~(s \in\{1,2\})$ are 
the collective parameters with $\aleph_{jl}^{(s)}$ and $\Omega_{jl}^{(s)}$ describing the mutual interactions among emitter-pairs $\{j,l\}$. For dipole-allowed transitions, 
for instance, one has $\aleph_{jl}^{(s)}=\sin{(\omega_{3s}r_{jl}/c)}/(\omega_{3s}r_{jl}/c)$ and  $\Omega_{jl}^{(s)}=-\cos{(\omega_{3s}r_{jl}/c)}/(\omega_{3s}r_{jl}/c)$ 
where we have averaged over all dipole orientations, whereas $r_{jl}=|\vec r_{j}-\vec r_{l}|$ are the inter-particle intervals between the $j$th and the $l$th emitters, 
respectively \cite{agarwal,gross,andre,garr,Ficek,rew,gxl,kaiser}. Further, $\omega_{\alpha\beta}$ with $\{\alpha,\beta\}\in\{1,2,3\}$ is the frequency of the 
$|\beta\rangle \leftrightarrow |\alpha\rangle$ atomic transition. $S^{(j)}_{\alpha\beta}=|\alpha\rangle_{jj}\langle\beta|$ represents the population of the state 
$|\alpha\rangle$ in the $j$-th atom, if $\alpha=\beta$, or the transition operator from $|\beta\rangle$ to $|\alpha\rangle$ of the $j$-th atom when $\alpha\neq\beta$. 
The atomic operators obey the commutation relations $\left[S^{(j)}_{\alpha\beta},S^{(l)}_{\beta'\alpha'}\right]=\delta_{jl}\left(\delta_{\beta\beta'}S^{(j)}_{\alpha\alpha'}- \delta_{\alpha\alpha'}S^{(j)}_{\beta'\beta} \right)$. Correspondingly, $\gamma_{1}$ and $\gamma_{2}$ are the single-atom spontaneous decay rates on 
$|3\rangle \to |1\rangle$ and $|3\rangle  \to |2\rangle$ atomic transitions.

\noindent{\it Results and feasible applications:} In the following, we shall use Eq.~(\ref{meq}) to investigate the collective dynamics of an initially excited ensemble of 
$\Lambda-$type emitters when $\gamma_{1} \gg \gamma_{2}$.

\noindent{\it Single-atom case:} For the sake of comparison, we first consider a single-atom case. The spontaneous decay law of an initially excited atom is given by the 
expression
\begin{eqnarray}
\langle S_{33}(t)\rangle =\langle S_{33}(0)\rangle \exp{[-2(\gamma_1+\gamma_2)t]},
\label{sat}
\end{eqnarray}
where $\langle S_{33}(0)\rangle$ denotes the initial population on the $|3\rangle$ level. This means that in the case of fully excited atom there is no way to influence 
the decay law of the upper state via applying a coherent laser field on the lower doublet levels. Furthermore, for a purely spontaneous decaying system the ratio of the 
lower states populations is: $\langle S_{11}(t)\rangle/\langle S_{22}(t)\rangle  = \gamma_{1}/\gamma_{2}$, i.e., these states will be spontaneously populated depending 
on the corresponding decay rates \cite{das}. Respectively, the spontaneous electromagnetic field intensities on these transitions are proportional to the population of the 
lower states during the spontaneous decay. Applying a coherent laser field on the lower doublet states, while the atom being initially on the upper excited state 
$|3\rangle$, the population among the lower energy-levels will oscillate after a while in the usual way.

\noindent {\it Multi-atom case:} In what follows, we shall see how these processes modify in the case of a collectively interacting atomic ensemble.
We shall continue by considering an ensemble of emitters with a higher density, i.e. $n\sim\lambda^{-3}_{2}$, such that the emitters on both involved 
transitions $|3\rangle\rightarrow |1\rangle$ and $|3\rangle\rightarrow |2\rangle$ interact collectively. Here $\lambda_{2}$ (or $\lambda_{1}$) is the 
corresponding wavelength on transition $|3\rangle\rightarrow |2\rangle$ ($|3\rangle\rightarrow |1\rangle$). Initially, the emitters are prepared in the 
excited state $|3\rangle$ and $\gamma_{1} \gg \gamma_{2}$. The dynamics of the cooperative decay on both transitions $|3\rangle\rightarrow |1\rangle$ 
and $|3\rangle\rightarrow |2\rangle$  is obtained with the help of master equation Eq.~(\ref{meq}) via decoupling of higher order atomic correlators - an 
approach valid for $N \gg 1$ \cite{andre}. Particularly, the equations for the population on the states 
$\langle S_{\alpha\alpha}\rangle/N=\sum^{N}_{j=1}\langle S_{\alpha\alpha}^{(j)}\rangle/N$, $\alpha \in \{1,2,3\}$, and the intensity of the superradiant 
emission $I_{\beta}\propto\langle S_{3\beta}S_{\beta3}\rangle/N^2=\sum^{N}_{j,l=1(j\neq l)}\langle S_{3\beta}^{(j)}S_{\beta3}^{(l)}\rangle/N^2$, 
$\beta \in\{1,2\}$, are governed by the number of collectively interacting emitters $N$, some geometrical factors $\{\mu_{1},\mu_{2}\}$ \cite{gross,andre}, 
the decay rates $\{\gamma_{1},\gamma_{2}\}$, and the Rabi frequency $\Omega$, respectively. To give some clarifications regarding the system of equations 
used to describe our sample, we present few terms in the equations of motion describing the population in the state $|1\rangle$ and the intensity on the 
$|3\rangle \to |1\rangle$ transition, namely, $(d/dt)\langle S_{11}\rangle$=$\cdots \sum_{l,k(l\neq k)}^{N}\gamma_{kl}^{(1)}\langle S_{31}^{(l)}S_{13}^{(k)}\rangle$ 
+ H.c., and $(d/dt)\langle S_{31}S_{13}\rangle$=$\cdots -\sum_{l,m,n(l\neq m\neq n)}^{N}\gamma_{ml}^{(1)}\langle S_{31}^{(l)}S_{13}^{(n)}(S_{11}^{(m)} - 
S_{33}^{(m)})\rangle - \sum_{l,m,n(l\neq m\neq n)}^{N}\gamma_{ml}^{(2)}\langle S_{32}^{(l)}S_{13}^{(n)}S_{21}^{(m)}\rangle$ + H.c., and so on. 
Here $\langle S_{11}\rangle$=$\sum^{N}_{j=1}\langle S^{(j)}_{11}\rangle$, whereas $\langle S_{31}S_{13}\rangle$=$\sum^{N}_{j\not=l}\langle S^{(j)}_{31}S^{(l)}_{13}\rangle$. One can observe that the equation of motion for a certain-order atomic correlator is represented through higher order ones. To obtain a closed system of equations 
we decoupled the three-particle correlators as follows: $\langle S_{31}^{(l)}S_{13}^{(n)}(S_{11}^{(m)}-S_{33}^{(m)})\rangle \approx \langle S_{31}^{(l)}S_{13}^{(n)}\rangle\langle(S_{11}^{(m)}-S_{33}^{(m)})\rangle$ and $\langle S_{32}^{(l)}S_{13}^{(n)}S_{21}^{(m)}\rangle \approx \langle S_{32}^{(l)}S_{13}^{(n)}\rangle \langle S_{21}^{(m)}\rangle$. The 'strategy' in decoupling procedure consists in trying to get a minimal system of equations 
of motion for a particular decoupling scheme, i.e., in our case the decoupling is applied on three-particle correlators (one can, for instance, start decoupling the 
four-particle correlators etc). At the end, we will arrive at a non-linear system of $12$ equations of motion, which are solved numerically. This method is widely 
used to characterize multiparticle ensembles \cite{andre}, and adequately describes collective intensities, populations, the fast decay etc., in the Dicke model or 
related systems/modifications.
\begin{figure}[t]
\includegraphics[width=4.1cm]{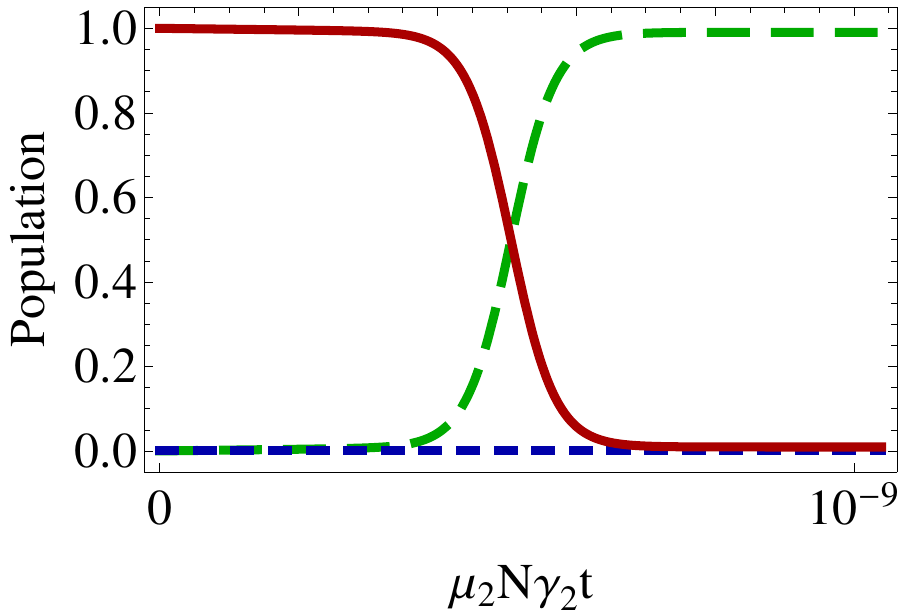}
\includegraphics[width=4.2cm]{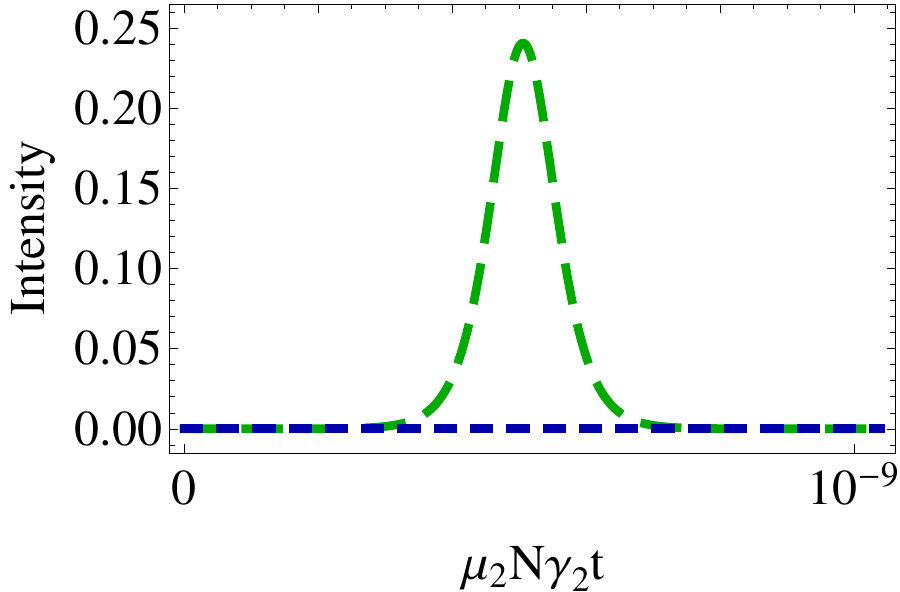}
\begin{picture}(0,0)
\put(-220,65){(a)}
\put(-20,65){(b)}
\end{picture}
\caption{\label{fig-2ab} (Color online) (a) The collective population on the state $|1\rangle$ (green long-dashed curve), the state 
$|2\rangle$ (blue short-dashed line) and the state $|3\rangle$ (red solid curve) as well as (b) the superradiant intensities 
on transitions $|3\rangle \to |1\rangle$ (green long-dashed curve) and $|3\rangle \to |2\rangle$ (blue short-dashed line) 
as a function of the scaled time $\mu_{2}N\gamma_{2}t$. Here $\Omega=0$, $\gamma_{2}/\gamma_{1}=10^{-8}$, 
$\mu_{2}/\mu_{1}=1/16$, $\mu_{2}=10^{-5}$ and $\langle S_{33}(0)\rangle = N=10^{7}$.}
\end{figure}
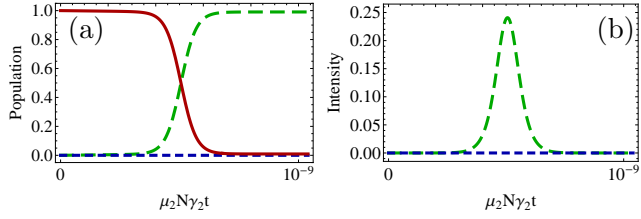

In the absence of the coherent driving, i.e. $\Omega=0$, the time-evolution of populations on the states $|1\rangle$, $|2\rangle$ and 
$|3\rangle$ as well as the collective intensities on the transitions $|3\rangle \rightarrow |1\rangle$ and $|3\rangle \rightarrow |2\rangle$ 
are presented in Fig.~\ref{fig-2ab}(a,b) when $\gamma_{1} \gg \gamma_{2}$. One can observe typical superradiant behaviors, that is, 
the population in the state $|3\rangle$ will cooperatively decay to the state $|1\rangle$ rapidly followed concomitantly by a superradiant 
pulse emission on transition $|3\rangle \rightarrow |1\rangle$ (see, respectively, the solid red curve in Fig.~\ref{fig-2ab}a and the green 
long-dashed lines in Fig.~\ref{fig-2ab}a and Fig.~\ref{fig-2ab}b). However, there is no superradiant emission on transition 
$|3\rangle \rightarrow |2\rangle$ (see the blue short-dashed lines in Fig.~\ref{fig-2ab}a and Fig.~\ref{fig-2ab}b). These behaviors can 
be well understood in the Dicke limit \cite{andre}. For an initially excited large atomic ensemble, i.e., $\langle S_{33}(0)\rangle=N$ and 
$N\gg 1$, one has
\begin{eqnarray}
\langle S_{11}(t)\rangle + 1 &=& \bigl(\langle S_{22}(t)\rangle + 1 \bigr)^{\gamma_{1}/\gamma_{2}}, ~~ {\rm~or} \nonumber \\
\langle S_{22}(t)\rangle + 1 &=& \bigl(\langle S_{11}(t)\rangle + 1 \bigr)^{\gamma_{2}/\gamma_{1}}. 
\label{popD}
\end{eqnarray}
It is easily to observe that if $\gamma_{1}=\gamma_{2}$ we always have $\langle S_{11}(t)\rangle = \langle S_{22}(t)\rangle$. 
For longer time-durations and when $\gamma_{1} \ll \gamma_{2}$ one has that $\langle S_{11}(t)\rangle \to 0$ whereas 
$\langle S_{22}(t)\rangle \to N$, and vice versa, i.e., for $\gamma_{1} \gg \gamma_{2}$, $\langle S_{22}(t)\rangle \to 0$ 
while $\langle S_{11}(t)\rangle \to N$.

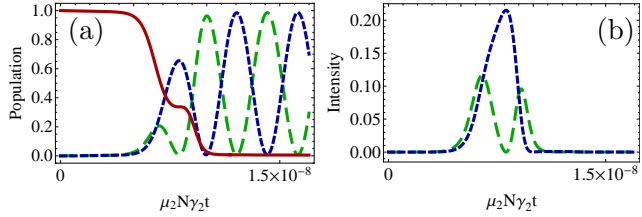
\begin{figure}[t]
\includegraphics[width=4.1cm]{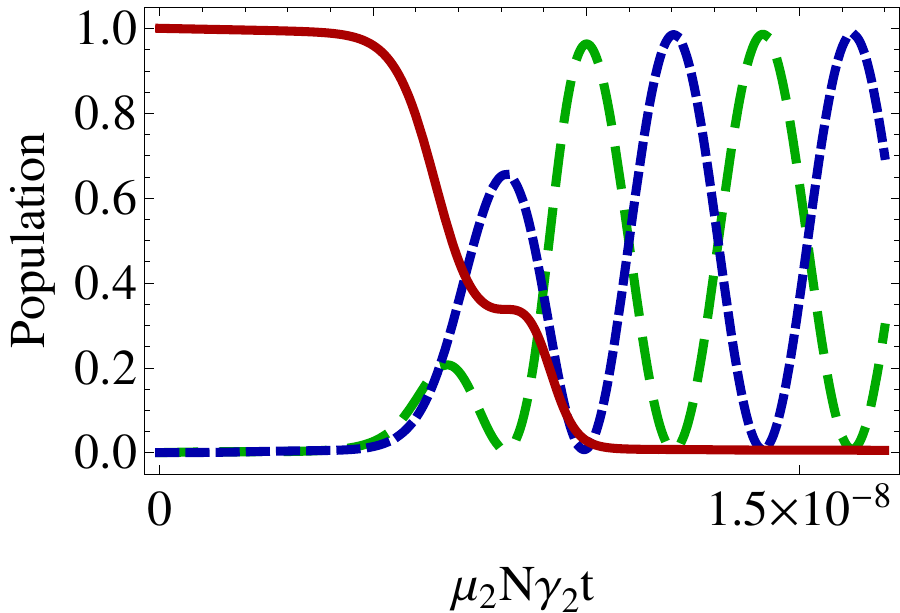}
\includegraphics[width=4.2cm]{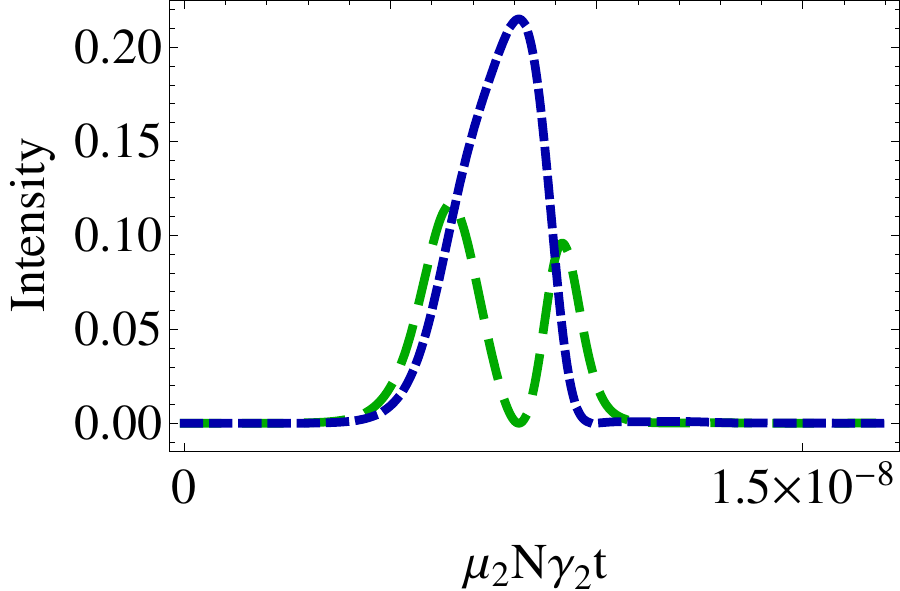}
\begin{picture}(0,0)
\put(-220,65){(a)}
\put(-20,65){(b)}
\end{picture}
\caption{\label{fig-3ab} (Color online) The same as in Fig.~\ref{fig-2ab} but for $\Omega/(\mu_{1}\gamma_{1}N)=0.47$.}
\end{figure}

Now we add a coherent laser field to couple the lower levels $|1\rangle \leftrightarrow |2\rangle$. This transition may be a dipole-forbidden  
one, therefore, it can be driven via two photon processes. If the Rabi frequency $\Omega$ is considerably smaller than the collective decay 
rates, i.e. $\Omega \ll \mu_{1}\gamma_{1}N$, there is only a very small amount of emitters decaying to the ground state $|2\rangle$ with 
a weak superradiant burst on transition $|3\rangle \rightarrow |2\rangle$, somehow similar to the picture described above. However, when 
the Rabi frequency is comparable but still smaller than the collective decay rate, i.e. $\Omega < \mu_{1}\gamma_{1}N$, the population 
dynamics is quite different from the case of smaller Rabi frequencies. Particularly, Figure~\ref{fig-3ab}(a,b) depicts the evolution of 
collective populations in the states $|1\rangle$, $|2\rangle$ and $|3\rangle$, as well as the intensities of the superradiant emissions for a 
particular value of the Rabi frequency, that is for $\Omega/(\mu_{1}\gamma_{1}N)=0.47$. Compared with the case $\Omega=0$ in 
Fig.~\ref{fig-2ab}(a), the population in the excited state $|3\rangle$ decreases to zero in a longer time and with a visible small plateau 
(see the red solid curve in Fig.~\ref{fig-3ab}a). On the other side, the population on the state $|2\rangle$ (blue short-dashed line in 
Fig.~\ref{fig-3ab}a) increases. 
The superradiance features behave accordingly. Remarkably, there is a strong superradiant pulse occurring 
on the much weaker transition $|3\rangle \rightarrow |2\rangle$ (see the blue short-dashed curve in Fig.~\ref{fig-3ab}b). Notice that the 
superradiant behaviors shown in Fig.~\ref{fig-3ab} differ from the ordinary superradiance in the sense that it is not quite determined by the 
fast population slope of the excited level since, in our case, the excited state population has almost a constant value when the superradiance 
peak occurs (for two-level emitters the superradiant intensity, $I$, is proportional to $I \propto - \partial \langle S_{z}(t)\rangle/\partial t$, 
where $\langle S_{z}(t)\rangle$ is the collective inversion operator). Although the most upper state population has a small plateau during a 
short time-interval (see the red solid curve in Fig.~\ref{fig-3ab}a), the population from the state $|1\rangle$ transfers, respectively, to 
$|2\rangle$ (see the long- and short-dashed curves in Fig.~\ref{fig-3ab}a), while the superradiance pulse achieves its maximum on 
$|3\rangle \leftrightarrow |2\rangle$ transition (see the blue short-dashed line in Fig.~\ref{fig-3ab}b). Furthermore, there are more atoms on the 
ground state $|2\rangle$ than in the excited one when the superradiant maximum takes place on the ultraslow transition (see Fig.~\ref{fig-3ab}). 
Again, this is distinct from standard superradiance features found in a two-level sample where the population is distributed equally when the 
superradiance peak occurs. Further, the intensity on the fast decaying transition vanishes as well as the population in the level $|1\rangle$ 
during the superradiant burst on the ultraslow transition (see Fig.~\ref{fig-3ab}). Additionally, due to the strong coupling between the states 
$|1\rangle$ and $|2\rangle$, the superradiant pulse on the fast transition $|3\rangle \rightarrow |1\rangle$ splits into two pulses (see the green 
long-dashed curve in Fig.~\ref{fig-3ab}b). At final stage, when the population on the state $|3\rangle$ reduces to zero, Rabi oscillations occur 
naturally among the states $|1\rangle$ and $|2\rangle$. These behaviors do not change much as long as $\Omega/(\mu_{1}\gamma_{1}N) \sim 1/2$.

The results described above can be physically explained in the semiclassical dressed-state picture. The corresponding eigenvectors due to 
laser-dressing of atoms on lower doublet levels can be written in terms of the bare states, namely,
\begin{eqnarray}
|\pm \rangle = \frac{1}{\sqrt{2}}\left(|2\rangle \pm |1\rangle \right).
 \label{dressed}
\end{eqnarray}
The energy difference between the two dressed states depends on the Rabi frequency of the driving coherent laser field $\Omega$. 
The population on the excited state $|3\rangle$ would decay to the two dressed states $|\pm\rangle$ which are a mixture of the bare 
states $|1\rangle$ and $|2\rangle$. Therefore, in the dressed-state picture, the intensity of the superradiant pulses on transitions 
$|3\rangle\rightarrow|1\rangle$ and $|3\rangle\rightarrow|2\rangle$ can be expressed as follows  
\begin{eqnarray}
I_{1} &\propto& \langle R_{3-}R_{-3}\rangle + \langle R_{3+}R_{+3}\rangle -\langle R_{3-}R_{+3}\rangle - \langle R_{3+}R_{-3}\rangle, 
\nonumber\\  
I_{2} &\propto& \langle R_{3-}R_{-3}\rangle + \langle R_{3+}R_{+3}\rangle + \langle R_{3-}R_{+3}\rangle + \langle R_{3+}R_{-3}\rangle. 
\nonumber \\ \label{int}
\end{eqnarray}
Here, $R_{3\pm}=\sum^{N}_{j=1}|3\rangle_{j}{}_{j}\langle \pm|$ ($R_{\pm 3}=\sum^{N}_{j=1}|\pm \rangle_{j}{}_{j}\langle 3|$) are the 
collective transition operators from the dressed states $|\pm\rangle \to |3\rangle$ ($|3\rangle \to |\pm\rangle$) of each emitter $j$. It follows 
from expressions~(\ref{int}) that the intensities of the superradiant pulses, 
while atoms decay from state $|3\rangle$ to states $|1\rangle$ and $|2\rangle$, include two parts: one part is the superradiance from state 
$|3\rangle$ to the dressed states $|\pm\rangle$ whereas the other part describes the contribution to the superradiant emission due to quantum 
coherences among the two decaying paths which are induced by the driving coherent source. When the Rabi frequency $\Omega$ is large, the 
emitters in the excited state $|3\rangle$ would decay via independent channels to the dressed states $|\pm\rangle$ because the cross-correlations 
among the two channels average out to zero. However, for smaller Rabi frequency, $\Omega < \mu_{1}\gamma_{1}N$, the two possible decaying 
pathways became indistinguishable such that the decay amplitudes from the excited state $|3\rangle \to |\pm \rangle$ interfere with each other. 
These collective decay-induced coherences may give rise to quantum interference between the two decaying paths. Actually, those decay-induced 
coherences lead to the constructive quantum interference on transition $|3\rangle \rightarrow |2\rangle$ whereas to destructive quantum interference 
on transition $|3\rangle \rightarrow |1\rangle$, respectively. That is why, for smaller Rabi frequencies, i.e. $\Omega=0.47\mu_{1}\gamma_{1}N$, 
a strong superradiant emission occurs on much weaker transition $|3\rangle \rightarrow |2\rangle$, while the superradiant pulse on transition 
$|3\rangle \rightarrow |1\rangle$ splits into two pulses. Respectively, the induced quantum coherences are responsible for population transfer 
among the lower doublet levels when the superradiant burst takes place, while the higher upper state population is almost constant. The whole 
cooperative process lasts during a time-period determined by the inverse of the faster collective decay rate. Notice that the cross-correlations 
among the two involved decay channels in the expressions (\ref{int}) vanish for a single-atom system (or many independent emitters), i.e., 
$R_{3-}R_{+3}=|3\rangle\langle-||+\rangle\langle 3|=0$, $R_{3+}R_{-3}=(R_{3-}R_{+3})^{\dagger}$, when $N=1$. Therefore, the effect 
described here is purely of the collective nature. Now we would like to compare the intensities emitted on ultraweak transition for independent 
emitters, $I_{2ind}$, or collectively interacting ones, $I_{2col}$. In the first case the intensity is: 
$I_{2ind} \sim \gamma_{2}N\langle S_{22}\rangle$, where $\langle S_{22}\rangle$ is the mean-value of single-atom population in the state $|2\rangle$. 
Taking into account that for independent or single-atom systems $\langle S_{22}\rangle/\langle S_{11}\rangle =\gamma_{2}/\gamma_{1}$ one has that 
$\langle S_{22}\rangle = (\gamma_{2}/\gamma_{1})/(1+\gamma_{2}/\gamma_{1})$. Thus, in this case, $I_{2ind}$=$\gamma_{2}N(\gamma_{2}/\gamma_{1})/(1+\gamma_{2}/\gamma_{1})$. For $\gamma_{2}/\gamma_{1}=10^{-8}$ and $N=10^{7}$, 
we have that $I_{2ind}=0.1\gamma_{2}$. For a collectively interacting ensemble, the peak intensity on ultraweak transition $|3\rangle \to |2\rangle$ 
can be estimated as: $I_{2col} \sim \gamma_{2}\mu_{2}N^{2}$.  For the same parameters as in Fig.~\ref{fig-3ab}(b), one has that: 
$I_{2}=20\gamma_{2}N$ which is significantly bigger than that for an independent atomic ensemble.

To create population inversions up to moderate x-rays frequencies may not be principially too hard because of available coherent light sources. 
Therefore, in these frequency ranges, our scheme may be applied for cooperative lasing or towards amplifying ultraslow atomic transitions like 
dipole-forbidden ones or due to parity violating effects \cite{par,weak,weakk}. Enhancing ultraweak transitions in quantum clock systems may 
be another option \cite{santra,sup_clock}. One may use a Lambda-type system containing ultranarrow optical transitions in alkaline-earth 
atoms (Sr, Yb, Ca, etc.), for instance \cite{conf}. For higher frequency effects it turns out that obtaining population inversion is quite 
challenging, although, one may proceed in the same vein as it was suggested in \cite{jetpL} to excite high lying energy levels in gamma 
diapason. 

\noindent{\it Summary:} We have investigated the superradiance effect occurring in a closely spaced $\Lambda-$type atomic  ensemble. 
The single-atom spontaneous decay rates to the lower doublet states are different. For an initially excited system, the superradiance 
phenomenon is taking place mainly on the transition possessing a higher spontaneous decay rate. We have found that when a coherent 
laser field is applied to the lower doublet states, the supperadiance is surprisingly enhanced on the much weaker atomic transition. This 
effect is identified with quantum interference effects among the decaying pathways which are induced by the presence of the coherent 
driving and it is not observed (i.e. emission enhancement due to quantum interferences) for a single-atom system or an independent 
atomic ensemble, respectively. Finally, the scheme works as well when $\gamma_{1} > \gamma_{2}$ or if $\omega_{31} \gg \omega_{32}$.

We have benefited from useful discussions with Christoph H. Keitel, Karen Z. Hatsagortsyan, Kilian Heeg, and Jonas Gunst. Also, we are grateful for the 
hospitality of the Theory Division of the Max Planck Institute for Nuclear Physics from Heidelberg, Germany. Furthermore, N.C. acknowledges the 
financial support from the National Natural Science Foundation of China (Grant No. 11404142), whereas M.M. acknowledges the financial support by 
the Max Planck Institute for Nuclear Physics as well as the Academy of Sciences of Moldova (grant No. 15.817.02.09F). 



\end{document}